\documentclass[9pt,twocolumn,twoside]{arxiv-styles/opticajnl}
\journal{arxiv} 

\setboolean{shortarticle}{true}

\usepackage{lineno}
\usepackage{soul}

\title{Five-dimensional single-shot fluorescence imaging using a polarized Fourier light-field microscope}

\author[*]{Oumeng Zhang}
\author[ ]{Changhuei Yang}

\affil[ ]{California Institute of Technology, Department of Electrical Engineering, Pasadena, California, 91125}

\affil[*]{ozhang@caltech.edu}

\begin{abstract}
Single-shot fluorescence imaging techniques have gained increasing interest in recent years due to their ability to rapidly capture complex biological data without the need for extensive scanning. In this letter, we introduce polarized Fourier light field microscopy (pFLFM), a novel fluorescence imaging technique that captures five-dimensional information (3D intensity and 2D polarization) in a single snapshot. This technique combines a polarization camera with an FLFM setup, significantly improving data acquisition efficiency. We experimentally validated the pFLFM system using a fluorescent Siemens star, demonstrating consistent resolution and an extended depth of field across various polarizations. Using the 5D imaging capabilities of pFLFM, we imaged plant roots and uncovered unique heterogeneities in cellulose fibril configurations across various root sections. These results not only highlight the potential of pFLFM in biological and environmental sciences, but also represent a significant advancement in the design of single-shot fluorescence imaging systems.
\end{abstract}

\setboolean{displaycopyright}{false} 

\begin{document}

\maketitle

Fluorescence imaging stands out as one of the most effective optical imaging techniques across various fields, including biological, biochemical, biomedical, and environmental studies, where it provides unparalleled insights due to its high specificity and sensitivity. Recent technological advances have increasingly focused on computational imaging techniques that enable the capture of multi-dimensional data in a single snapshot. Traditional volumetric fluorescence microscopy methods, such as confocal or light sheet microscopy, typically require scanning to reconstruct 3D structures. This not only complicates the imaging system with the need for multiple moving parts but also extends the time required for data acquisition. In contrast, single-shot volumetric fluorescence (SVF) imaging techniques \cite{boominathan_recent_2022} offer a promising alternative by encoding 3D information into 2D measurements through the design of the system's point spread function (PSF). Subsequently, a reconstruction algorithm processes this data to extract the 3D information. In addition to video-rate acquisition capabilities, the simplicity of these setups in some of the latest SVF imaging techniques enable miniaturization \cite{adams_single-frame_2017,yanny_miniscope3d_2020,xue_single-shot_2020}, significantly expanding their applicability in \textit{in vivo} applications.

One of the approaches in SVF imaging is Fourier light field microscopy (FLFM) \cite{guo_fourier_2019,han_3d_2022}. In contrast to traditional light field microscopy \cite{levoy_light_2006,broxton_wave_2013,skocek_high-speed_2018}, which places a microlens array (MLA) at the image plane to capture light field information, FLFM positions the MLA at the pupil plane of the imaging system. In this configuration, each lenslet in the MLA captures the object from a slightly different angle, creating perspective images at the focal plane of the MLA, all of which are recorded by a single camera. This arrangement mirrors the mechanism of binocular vision, enabling 3D reconstruction of the object. Recent advancements have further improved FLFM's capabilities. For example, using lenslets with varying focal lengths has extended the depth of field by an order of magnitude \cite{linda_liu_fourier_2020}. Additionally, replacing traditional sensors with event cameras allows for ultrafast 3D imaging \cite{guo_eventlfm_2024}.

Another significant direction in the advancement of fluorescence imaging technology involves measuring properties beyond the intensity distribution, such as fluorescence lifetime and emission spectrum. Among these, fluorescence polarization is particularly significant because it provides insights into the orientation of fluorophores, which can reveal details about the local biochemical and biophysical environment \cite{brasselet_polarization_2023,zhang_single-molecule_2024}. Briefly, fluorescent molecules can be modeled as oscillating dipoles that emit polarized light, with the emission polarization depending on the orientation of the dipole \cite{Axelrod2012,Backer2014}. When fluorophores are randomly oriented within a sample, the resulting fluorescence emission is typically isotropic, showing no preferred polarization direction. In contrast, when fluorophores are more orderly arranged within the local environment, a strong polarization preference in the emitted fluorescence is observed, indicating the structured alignment of these fluorophores. This property has been used to explore various biological phenomena, such as chemical heterogeneities within lipid membranes \cite{Lu2020,Zhang2022,zhang_six-dimensional_2023}, the growth and decay of amyloid fibrils \cite{sun_single-molecule_2024}, and conformational changes in DNA structures \cite{backer_single-molecule_2019}. 

In this letter, we introduce polarized Fourier light field microscopy (pFLFM), a fluorescence imaging technique capable of capturing five-dimensional information (3D intensity and 2D polarization) in a single snapshot. This method integrates a polarization camera with an FLFM setup. We experimentally validated the pFLFM system using a fluorescent Siemens star. Leveraging its 5D imaging capability, we have applied pFLFM to image plant roots, revealing distinct cellulose fibril configurations within various regions of the roots. These results demonstrate the feasibility of 5D single-shot imaging and underscore its biological relevance by providing unprecedented insights into the structural organization and orientation of cellular components.

\begin{figure}[t!]
    \centering
    \includegraphics{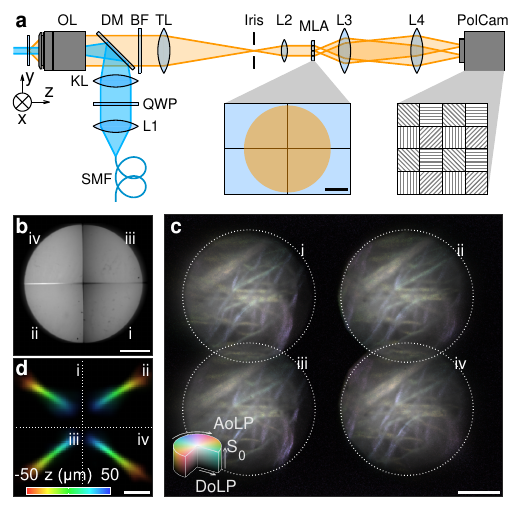}
    \caption{(a) Schematic of the pFLFM system. SMF: single mode fiber; QWP: quarter-wave plate; KL: K\"{o}hler lens; OL: objective lens; DM: dichroic mirror; BF: bandpass filter; TL: tube lens; L1-L4: lenses; MLA: microlens array; PolCam: polarization camera. Insets show the alignment of the pupil relative to the lenslets in the MLA and the configuration of polarized pixels in the polarization camera. (b) Image of the pupil and the MLA. (c) Representative image of wheat root captured with the pFLFM system. The field of view in channels (i)-(iv) is set by the iris. Images are color-coded according to the angle of linear polarization (AoLP), degree of linear polarization (DoLP), and Stokes parameter $S_0$ (intensity). (d) PSF color-coded with depth $z$ in channels (i)-(iv), which correspond to four sections of the pupil. Scale bar: 500 \textmu{}m in (b), 200 \textmu{}m in (c), and 20 \textmu{}m in (d).}
    \label{fig:schematic}
\end{figure}
The imaging system is shown in Figure \ref{fig:schematic}a. 
A 488-nm laser (Thorlabs LP488-SF20G) coupled to a single-mode fiber (SMF) was modulated by lens L1, a quarter-wave plate (QWP) and a K\"{o}hler lens (KL).
After reflection from a dichroic mirror (DM, Semrock Di01-R488/561), the resulting illumination at the sample was uniform and circularly polarized. 
A 20x, 0.4-NA objective lens (Olympus PLN20X) and a tube lens (TL, $f=150$ mm) were used to create an intermediate image plane.
An iris was positioned at this plane to adjust the field of view. 
The captured fluorescence was filtered using a bandpass filter (BF, Semrock FF01-523/610).
Lens L2 ($f=40$ mm) projected the pupil plane onto a microlens array (MLA, Thorlabs MLA1M), where the pupil was demagnified to align precisely with four lenslets of the MLA (Figure \ref{fig:schematic}b). A 4f system (L3, $f=30$ mm and L4, $f=80$ mm) then magnified and projected the images from the MLA onto a polarization camera (PolCam, The Imaging Source DZK 33UX250).

The raw image captured by the polarization camera (Figure \ref{fig:schematic}c) consists of four sub-images corresponding to pixel intensities along 0°, 45°, 90°, and 135° polarization axes. These polarized images can be used to compute the three Stokes parameters $S_{0,1,2}$, which can be further mapped to intensity (equal to $S_0$), angle of linear polarization (AoLP), and degree of linear polarization (DoLP). To visualize these parameters simultaneously, we apply a colormap where color represents the AoLP, saturation represents the DoLP, and brightness represents the intensity \cite{bruggeman_polcam_2024}. Specifically, white denotes isotropic polarization, a highly saturated color indicates strong linear polarization along a specific direction, and black represents low intensity. This colormap is also used for visualizing the reconstruction results.

\begin{figure}[t!]
    \centering
    \includegraphics{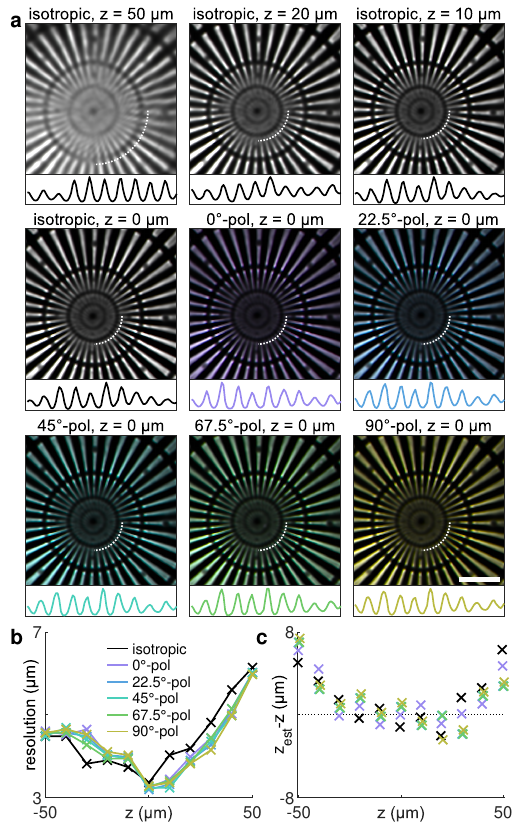}
    \caption{(a) Representative reconstruction of a fluorescent Siemens star under varying defocus and fluorescence polarizations. Scale bar: 50 \textmu{}m. (b) Lateral resolution and (c) estimation bias ($z_\mathrm{est}-z$) as functions of defocus ($z$) for isotropic fluorescence and linearly-polarized fluorescence at different angles.}
    \label{fig:star}
\end{figure}

\begin{figure*}[ht!]
    \centering
    \includegraphics{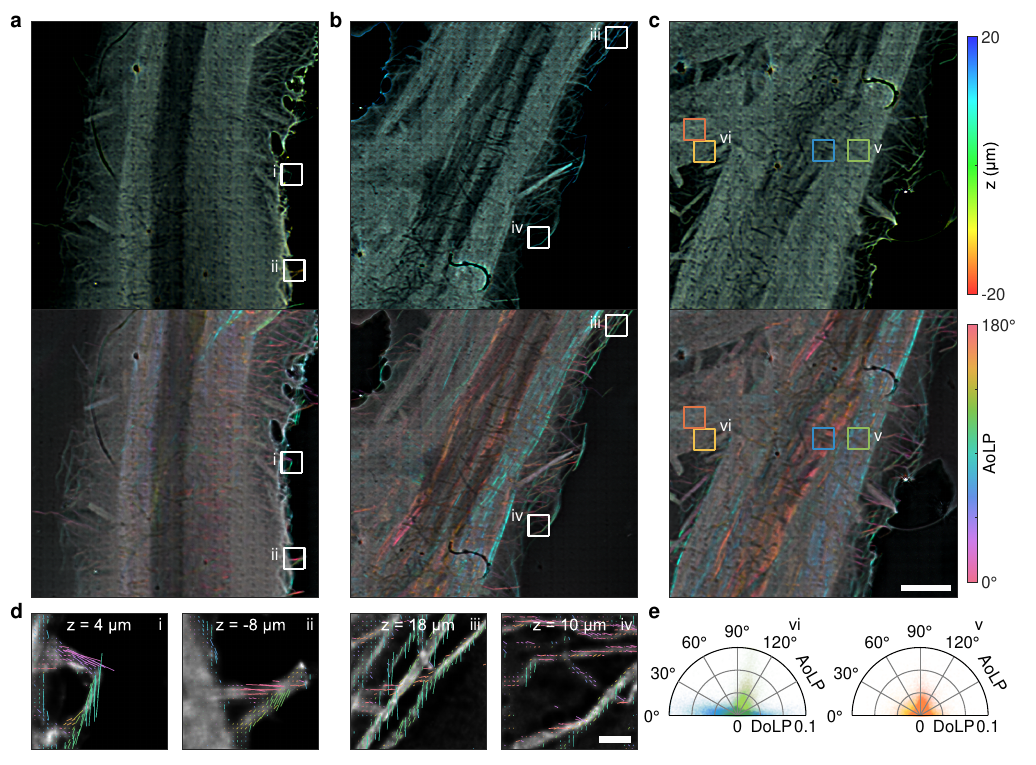}
    \caption{(a-c) Reconstructions of wheat root sections, color-coded by (top) height and (bottom) polarization. (d) Zoomed-in views of the boxed regions in (a,b) at the sharpest focal depths. Lines are oriented and color-coded according to the AoLP and their lengths are proportional to the DoLP. (e) Distribution of AoLP and DoLP for the zoomed-in regions in (c). Scale bar: 500 \textmu{}m in (a-c); 50 \textmu{}m in (d).}
    \label{fig:root}
\end{figure*}

Considering the aberrations introduced by the MLA, we opted to use experimentally captured PSFs instead of theoretical PSFs for image reconstruction (Figure \ref{fig:schematic}d). These PSFs were obtained by scanning fluorescent beads across a range from -50 to +50 \textmu{}m. The z-dependent PSFs show characteristic features typical of an FLFM system, where focal spots in each channel exhibit lateral shifts in different directions corresponding to the lenslets' positions on the pupil.

To reconstruct the 3D volume along with the AoLP and DoLP, raw images are first segmented into 0°-, 45°-, 90°-, and 135°-polarized sub-images. These sub-images, along with the experimental PSF, are processed using a modified Richardson-Lucy deconvolution algorithm \cite{dellacqua_model-based_2007,guo_fourier_2019} to reconstruct 3D volumes for each of the four polarizations. These volumes are then combined to produce the volumetric AoLP, DoLP, and intensity.

To demonstrate the pFLFM system, we first evaluate its resolution by imaging a fluorescent Siemens star target with both isotropic fluorescence emission and polarized emission (Figure \ref{fig:star}). For polarized emission, a linear polarizer was placed between lenses L3 and L4 to simulate fluorophores aligned along a specific direction. It is important to emphasize that although the PSF for polarized fluorescence from dipole-like emitters differs from that of isotropic emitters such as fluorescent beads, the use of a lower NA (0.4) objective lens significantly reduces these differences. Specifically, an isotropic emitter can be viewed as a mixture of $x$-, $y$-, and $z$-oriented dipole emitters. When the emitted fluorescence passes through an $x$-oriented linear polarizer, 95.94\% of the detected intensity originates from the $x$-oriented dipole, while only 4.03\% and 0.03\% come from the $z$- and $y$-oriented components, respectively. This confirms that the use of a linear polarizer effectively mimics real fluorescence emission from dipole-like emitters with an in-($xy$)plane orientation.

Figure \ref{fig:star}a shows representative reconstruction results and line profiles near the resolution limit. We first examine results from isotropic fluorescence. When compared to the reconstruction of an in-focus Siemens star, a defocus of 10-\textmu{}m shows no resolution degradation, whereas a 20-\textmu{}m defocus leads to slightly reduced resolution. At $z=50$ \textmu{}m, there is evident blur induced by the defocus. For reconstructions of an in-focus Siemens star with linearly polarized fluorescence, the results show polarization measurements that align with the ground truth. Further, the resolution remains consistent with that observed in reconstructions from isotropic fluorescence. This consistency confirms the robustness of the imaging system in handling varied fluorescence polarizations.

Quantitative analysis (Figure \ref{fig:star}b) further supports our observations. The in-focus resolution is 3.4 \textmu{}m for isotropic fluorescence and 3.2, 3.2, 3.2, 3.3, and 3.3 \textmu{}m for linearly polarized fluorescence at 0°, 22.5°, 45°, 67.5°, and 90°, respectively. These values are slightly larger than the sampling limit of 2.6 \textmu{}m in our system (the pixel size is 1.3 \textmu{}m at the object plane). Notably, the resolution remains below 4.2 \textmu{}m over a depth of field of $\sim$50 \textmu{}m for all polarizations. This depth of field significantly exceeds the specification of the objective lens (1.72 \textmu{}m). Additionally, we evaluated the depth accuracy (Figure \ref{fig:star}c) by comparing the difference between the estimated depth ($z_\mathrm{est}$) and the ground truth ($z$). Within $|z|\leq30$ \textmu{}m, the system exhibits high estimation accuracy, with an average absolute bias ($|z_\mathrm{est}-z|$) of 0.9 \textmu{}m and a maximum of 2.4 \textmu{}m across all polarizations. However, beyond this range, the absolute bias increased, averaging 5.2 \textmu{}m with a maximum reaching 7.4 \textmu{}m for $|z|=50$ \textmu{}m.

To demonstrate one of the biological applications of pFLFM, we imaged plant roots stained with Congo Red, a widely used fluorescent dye that binds to 1,4-glucans \cite{wood_specificity_1980}; its emission dipole aligns with the cellulose fibrils in plant cells, exhibiting strong polarization features. This property makes it particularly effective for revealing the orientation of cellulosic fibers in various plant species \cite{verbelen_polarization_2000,thomas_detection_2017,piccinini_imaging_2024}. In this study, we focused on common wheat (\textit{Triticum aestivum}). The roots were treated with 1M KOH overnight, then rinsed with 0.1M HCl before being incubated in a 1\% Congo Red solution in water overnight. These prepared samples were then imaged using pFLFM.

The reconstructions of root sections are shown in Figure \ref{fig:root}a-c. Using the 3D capabilities of pFLFM, we can observe variations in the heights of root hair structures, with representative zoomed-in regions at $z = 4$, -8, 18, and 10 \textmu{}m, respectively (Figure \ref{fig:root}d). Polarization measurements in the root hair region indicate that the direction of polarization is predominantly parallel to the orientation of the root hairs (Figure \ref{fig:root}d), suggesting that the cellulose fibrils are aligned with these structures. Notably, in the primary root, distinct polarization features are observed between the center and outer regions; the polarization in the outer region is mostly parallel to the root, whereas in the center, it appears perpendicular (Figure \ref{fig:root}e(v)). This suggests different cellulose fibril organizations within these two areas. A similar variation in polarization is also observed in a secondary root branch, although the differences appear more subtle (Figure \ref{fig:root}e(vi)).

In summary, we introduce pFLFM, a 5D single-shot fluorescence imaging technique that combines FLFM, a widely used SVF method, with polarized detection. This approach not only captures the 3D fluorescence intensity distribution but also measures the angle and degree of linear polarization of the fluorescence in a single snapshot. The system maintains consistent resolution across various polarizations and extends the depth of field significantly compared to that of conventional 2D widefield microscopy. Its 5D imaging capability has enabled us to reveal detailed structural heterogeneities within wheat root sections.
These results provide deeper insights into plant cell architecture, which demonstrates the potential of pFLFM to contribute to the fields of biological and environmental sciences.

While our work has successfully demonstrated the capabilities of 5D single-shot fluorescence imaging, there remains significant potential to further improve this technology. For instance, rather than using an MLA, developing a custom polarization and phase pattern optimized for 5D imaging, e.g., using a metasurface, could be paired with the polarization sensor to improve the image quality. Additionally, our current setup does not consider the polarization mixture caused by the $z$-oriented component of the fluorophore's orientation, an effect that could become significant with the use of a high NA objective lens. Exploring how a more sophisticated vectorial emission model \cite{Axelrod2012,Backer2014} impacts the measurements, such as potential intensity reconstruction artifacts, is crucial for improving the resolution of this technique. Another promising direction involves addressing the potential overlap between sub-images using a more advanced algorithm \cite{yang_wide-field_2024}, which could expand the field of view and further improve the spatio-temporal throughput. We hope that this work will spark interest and drive further research in multi-dimensional single-shot fluorescence imaging.

\begin{backmatter}
\bmsection{Funding} The work reported in this paper is supported by the Resnick Sustainability Institute. 

\bmsection{Acknowledgment} We thank Reinaldo Alcalde for providing common wheat roots used in this study.

\end{backmatter}

\bibliography{references}


\end{document}